\documentstyle[aaspp4,epsf]{article}

\begin{document}
\lefthead{White, Scott \& Pierpaoli}
\righthead{Boomerang returns unexpectedly}

\hyphenation{an-is-o-tro-pies}
\hyphenation{an-is-o-tro-py}

\title{Boomerang returns unexpectedly\footnote{A paper about a boomerang
by an Australian and his mates.}}

\author{Martin White}
\affil{Harvard-Smithsonian Center for Astrophysics,\\
60 Garden Street, Cambridge, MA 02138}
\author{Douglas Scott and Elena Pierpaoli}
\affil{Department of Physics \& Astronomy,\\
University of British Columbia, Vancouver, BC, V6T 1Z1}

\authoremail{mwhite@cfa.harvard.edu}
\authoremail{dscott@astro.ubc.ca}
\authoremail{elena@astro.ubc.ca}

\begin{abstract}
\noindent
\rightskip=0pt
Experimental study of the anisotropy in the cosmic microwave background (CMB)
is gathering momentum.  The eagerly awaited {\sl Boomerang\/} results have
lived up to expectations.  They provide convincing evidence in favor of the
standard paradigm: the Universe is close to flat and with primordial
fluctuations which are redolent of inflation.  Further scrutiny reveals
something even more exciting however -- two hints that there may be some
unforeseen physical effects.
Firstly the primary acoustic peak appears at slightly larger scales than
expected.  Although this may be explicable through a combination of mundane
effects, we suggest it is also prudent to consider the possibility that the
Universe might be marginally closed.  The other hint is provided by a second
peak which appears less prominent than expected.  This may indicate one of a
number of possibilities, including increased damping length or tilted initial
conditions, but also breaking of coherence or features in the initial
power spectrum.  Further data should test whether the current concordance
model needs only to be tweaked, or to be enhanced in some fundamental way.
\end{abstract}

\rightskip=0pt
\section{Introduction}

The study of the Cosmic Microwave Background (CMB) anisotropy holds the
promise of answering many of our fundamental questions about the Universe
and the origin of the large-scale structure
(see e.g.~Bond~\cite{Bond}; Bennett, Turner \& White~\cite{BenTurWhi};
Lawrence, Scott \& White~\cite{Romans}).
The development of CMB research can be split into 5 main phases.
Firstly, the mere existence of the CMB showed that the early Universe was hot
and dense.  Secondly, the blackbody nature of the CMB spectrum and its
isotropic distribution implied that the Universe is approximately homogeneous
on large scales.
The third step came with the detection of anisotropies, confirming that
structure grew through gravitational instability.
Now we are entering the fourth stage, where the basic cosmological paradigm
is defined.
The recently released {\sl Boomerang\/} data (de Bernardis et al.~\cite{BOOM})
provide support for a model with adiabatic initial conditions and a Universe
with approximately flat geometry.
The fact that our theories are holding up so well gives us further reason to
believe that the CMB can be used as a precision cosmological tool.
With the imminent launch of {\sl MAP\/}, we are on the verge of the fifth
phase, which involves determining the precise values of the fundamental
cosmological parameters to figure out exactly what kind of Universe we live in.

Most of the unmined cosmological information available from the CMB anisotropy
is encoded in the acoustic signatures, the series of peaks and troughs in the
spectrum at subdegree scales, which we are only now beginning to probe
experimentally.
Because the properties of the photon-baryon oscillations are determined by the
background, while the driving force is described by the model for the
perturbations, the acoustic signatures provide a unique opportunity to probe
both the background cosmology and the model for structure formation.
For example the position of the first peak, or indeed any other feature,
provides a measure of the angular diameter distance to last scattering.
The relative heights of the peaks provide information about the baryon
`drag' on the photons and thus the baryon-to-photon ratio.
The relative peak locations provide information on the perturbations as they
crossed the horizon and thus indirectly on the mechanism for their production
(see e.g.~Hu, Sugiyama \& Silk~\cite{HuSugSil}).

In the last year or so there have been several new CMB data sets which have
begun to reveal the structure contained in the acoustic peaks
(see e.g.~Lineweaver~\cite{Charley}; Dodelson \& Knox~\cite{DodKno};
Melchiorri et al.~\cite{Mel}; Pierpaoli, Scott \& White~\cite{Science};
Efstathiou~\cite{Efs};
Tegmark \& Zaldarriaga~\cite{TegZal}; Lahav et al.~\cite{LahBriHobLasSod};
Le Dour et al.~\cite{LeDour}~for analyses of these data).
Now with the first estimate of the power spectrum from a sub-set of the
Antarctic flight of the {\sl Boomerang\/} experiment we are entering a whole
new regime of precision.  There are 3 striking things about this new power
spectrum estimate.  Firstly, and most importantly, it corroborates the basic
picture of cosmological structure formation -- the shape is a confirmation
of flat models of the sort inspired by inflation, dominated by a cosmological
constant, as has become the standard paradigm.  Secondly, however, the
position of the first peak appears at slightly larger angular scales than
might have been expected.  And lastly, another possibility for something
unexpected comes through a hint that the second peak may not be as pronounced
as most models would predict.  We will make some general comments about the
existence of the first acoustic peak, and then in the rest of this
paper we focus on these latter two surprising features of the new
data (see also Hu~\cite{Hu}).

\section{A distinct acoustic peak}

The presence of a narrow, well defined peak in the angular power spectrum
(for which earlier evidence existed:
Dodelson \& Knox~\cite{DodKno};
Melchiorri et al.~\cite{Mel};
Pierpaoli, Scott \& White~\cite{Science})
has important consequences.
Primarily it shows that the theoretically expected acoustic peaks are in fact
present in nature!
A well defined narrow peak implies that whatever caused the fluctuations did
so at very early times rather than actively driving the photon-baryon fluid
at recombination (as would happen in models based on topological defects,
for example).  By the time the anisotropies formed at $z\,{\sim}\,10^3$, the
growing mode of the perturbations was dominant.  The width of the peak is then
essentially a measure of the inertia of the baryon-photon fluid at last
scattering.  The finite inertia of the fluid provides an upper limit to how
fast it can respond to any impulse, and thus a minimum width to any feature
in the angular power spectrum.

We show in Fig.~\ref{fig:bp} the {\sl Boomerang\/} data from
de Bernardis et al.~(\cite{BOOM}) along with some theoretical models and a
compilation of older data from Pierpaoli et al.~(\cite{Science}).
In comparison with the older data, the {\sl Boomerang\/} data appear lower
around the first peak.  Note however that the Pierpaoli et al.~(\cite{Science})
points are somewhat anti-correlated with their nearest neighbors.
Thus the disagreement is not quite as large as it appears.
Moreover, any remaining discrepancy is consistent with the ${\sim}\,10$\%
calibration uncertainty between experiments.
In other words, on a power spectrum plot, one is allowed to shift the power
spectrum estimates from individual data sets by as much as 20\% vertically
relative to each other.
It appears then that the {\sl Boomerang\/} data set has a lower overall
calibration (as did the data from the {\sl Boomerang\/} test flight,
Mauskopf et al.~\cite{BOOM97}) than some of the earlier experiments.
However, calibration issues aside, there is remarkable agreement about the
shape of the power spectrum around multipole $\ell\,{\sim}\,200$.

It is a dramatic verification of the simplest cosmological models that the
existence of such a peak, made at least as early as 1970
(Peebles \& Yu~\cite{PeeYu}, where the baryon power spectrum was plotted,
although the peaks were implicitly there in the photon power spectrum also)
and more explicitly by the mid-1970s
(e.g.~Doroshkevich, Sunyaev \& Zel'dovich~\cite{DSZ}, where peaks are
shown in the CMB $k$-space power spectrum), has been confirmed by
experiment.  This sort of clear test of theoretical ideas is quite
uncommon in cosmology!

\section{First peak position}

The {\sl Boomerang\/} data show a peak which lies at lower $\ell$ than the
canonical value for a flat universe, which is $\ell\,{\simeq}\,220$.
The first explanation
for this would be that we live in a closed universe; however it is interesting
to ask what other options exist.
Firstly, the peak position comes from a quadratic fit to the data between
$\ell=50$ and 300 (de Bernardis et al.~\cite{BOOM}).  Thus at least some of
the constraint pushing the peak to lower $\ell$ is in the rapid decrease
in power beyond the peak.  We have checked that the peak position is unaffected
by the precise functional form used, but it remains true that the low effective
$\ell_{\rm peak}$ may be explained partly by the same physics that makes the
second peak lower than expected.  Nevertheless, Fig.~1 shows clearly that the
$\Lambda$-dominated `concordance model' (Ostriker \& Steinhardt~\cite{OstSte})
does not give a good fit around the peak.

As shown in Hu \& White~(\cite{NewTest}; \cite{Signatures}), using the first
peak to measure the angular diameter distance to last scattering can be a
subtle business.  Fortunately the positions of the peaks are sensitive to few
of the many other cosmological parameters that alter the anisotropy
spectrum.  If the baryon density is constrained to satisfy big-bang
nucleosynthesis, then it introduces a negligible uncertainty on the peak
positions.  Higher order effects do not change the peak positions.
Of course, any effect on the power spectrum which reduces small-scale power
will also shift the peak a little to the left, but this is essentially
negligible for reasonable parameters (e.g.~for tilt $0.7\le n\le 1.3$).

The major effect, then, is the angular diameter distance to last scattering and
the physical matter density $\propto\Omega_{\rm M}h^2$.
If we hold the distance to the last scattering surface fixed a low matter
density universe, which has last scattering closer to the radiation dominated
epoch, has a peak broadened and shifted leftwards.  This effect
is small, however, and is usually overcome by the cosmological dependence of
the distance to last scattering.
For a flat universe with fixed $\Omega_{\rm B}h^2$, the distance to the last
scattering surface is a function of $\Omega_{\rm M}$ and $h$, being shorter
for high $\Omega_{\rm M}$ or $h$.  Thus the $\ell$ of the first peak
decreases slightly with increasing $\Omega_{\rm M}$ or $h$.
We show this in Fig.~\ref{fig:peak}, where we plot the $\ell$ of the first
peak as a function of $\Omega_{\rm M}$ for flat models.
For reasonable cosmological parameters the first peak can be as low as 210
even in a flat universe.  For $\ell_{\rm peak}\,{<}\,210$,
one starts violating other cosmological constraints.

Thus we have suggestive evidence that the Universe may be spatially closed.
This runs counter to the observationally motivated convention of the last
several years to consider only open models, and the theoretically motivated
desire to extend that precisely as far as flat.
Inflationary models certainly exist (e.g.~Linde~\cite{Lin}),
in which a closed universe
is created `from nothing' (Zel'dovich \& Grishchuk~\cite{ZelGri}).
Historically there has been much interest in closed universes (see
Bj\"{o}rnsson \& Gudmundsson~\cite{BjoGud},
White \& Scott~\cite{WhyNot}, and references therein), since the
spatial surfaces are compact (Wheeler~\cite{Whe}; Hawking~\cite{Haw}) and the
total energy, charge and angular momentum are zero
(Landau \& Lifshitz~\cite{LanLif}).
This has appealing properties of finiteness and flux conservation for formal
studies, and hence has been preferred by various authors on grounds that are
essentially philosophical, or at least mathematical
(for an interesting historical discussion see
Misner, Thorne \& Wheeler~\cite{MTW}, \S 21.12 and \S 27.1).
In other words, faced with the choice of $\Omega_{\rm tot}=1+\epsilon$ or
$1-\epsilon$, there may be reasons to choose the former.  Whether or not
the Universe turns out to be spatially closed, open or precisely flat
remains an empirical question.  The suggestion that some closed models may
provide a good fit to current data means {\it at the very least\/}
that one should
be democratic to both sides of the curvature likelihood function
when carrying out multi-parameter cosmological explorations.

\section{Second peak height}

\begin{table}
\begin{center}
\begin{tabular}{l|cccc}
& \multicolumn{4}{|c}{$\Omega_{\rm B}$} \\
$\Omega_{\rm M}$ & 0.03 & 0.04 & 0.05 & 0.06 \\ \hline
0.2 & 1.81 & 2.02 & 2.32 & 2.62\\
0.3 & 1.70 & 1.97 & 2.27 & 2.62\\
0.4 & 1.64 & 1.92 & 2.26 & 2.64\\
0.5 & 1.61 & 1.90 & 2.26 & 2.67\\
\end{tabular}
\end{center}
\caption{The ratio of the height of the first peak to the second peak, for
a model with $h\,{=}\,0.7$ and $n\,{=}\,1$.
This ratio is fixed by the physics at
recombination and the initial perturbation spectrum, thus it depends on
$\Omega_{\rm M}h^2$, $\Omega_{\rm B}h^2$, and $n$.}
\label{tab:ratio}
\end{table}

\begin{figure}
\begin{center}
\leavevmode
\epsfysize=10cm \epsfbox{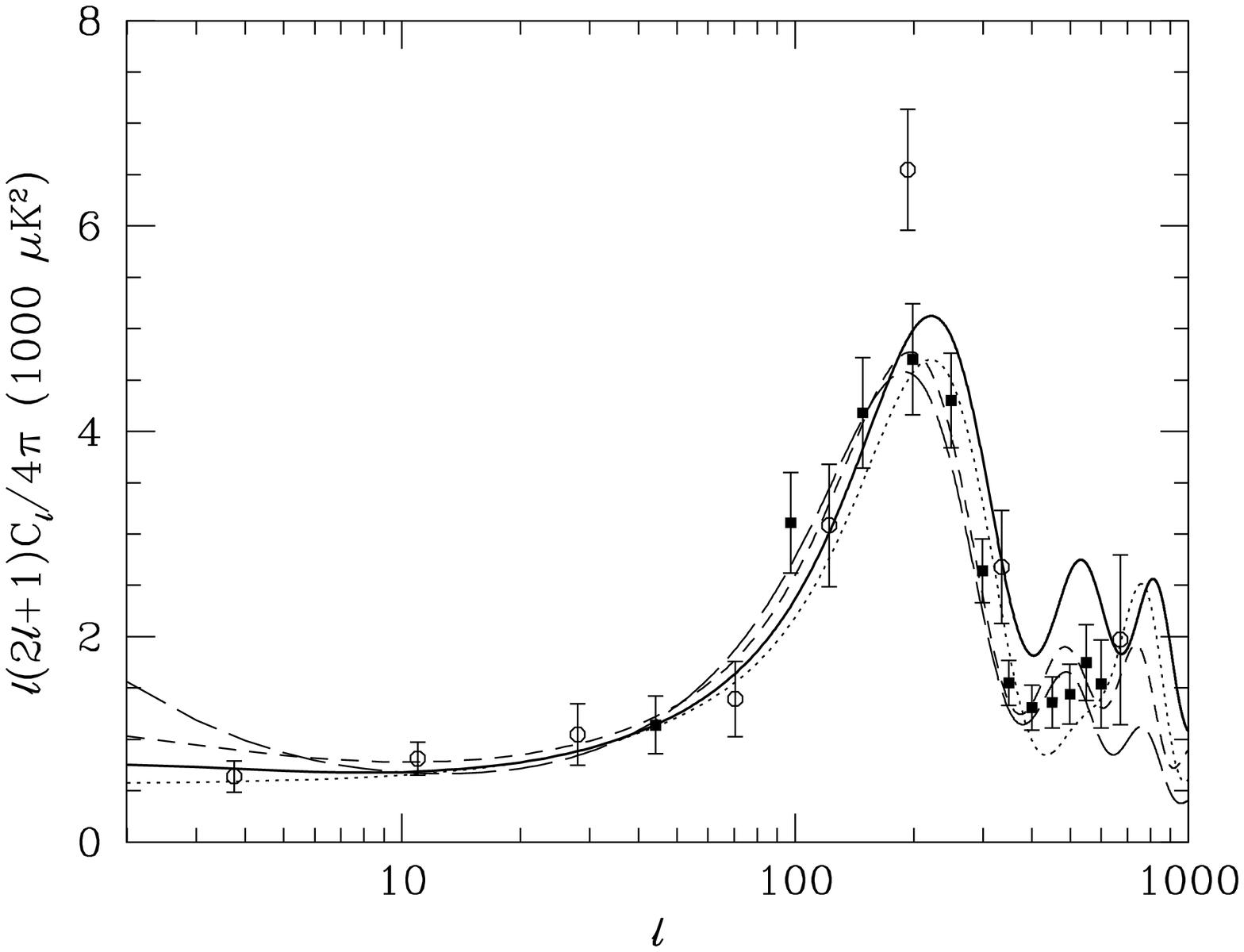}
\end{center}
\caption{A comparison of the {\sl Boomerang\/} data (solid squares) with an
earlier compilation (Pierpaoli et al.~\protect\cite{Science}; open circles)
and some theoretical models.  The solid line is the `standard' $\Lambda$CDM
model of Ostriker \& Steinhardt~(\protect\cite{OstSte}).  The dashed line
is an example of a model that has been tweaked to provide a better fit to the
{\sl Boomerang\/} data: a slightly closed, high baryon, tilted $\Lambda$CDM
model with $\Omega_{\rm M}=0.4$, $\Omega_\Lambda=0.7$, $h=0.6$,
$\Omega_Bh^2=0.025$ and $n=0.9$.  The dotted line is a critical density model
with a high baryon fraction $\Omega_{\rm B}=0.1$ while the long-dashed line
is the decaying neutrino model discussed in the text.}
\label{fig:bp}
\end{figure}

\begin{figure}
\begin{center}
\leavevmode
\epsfysize=10cm \epsfbox{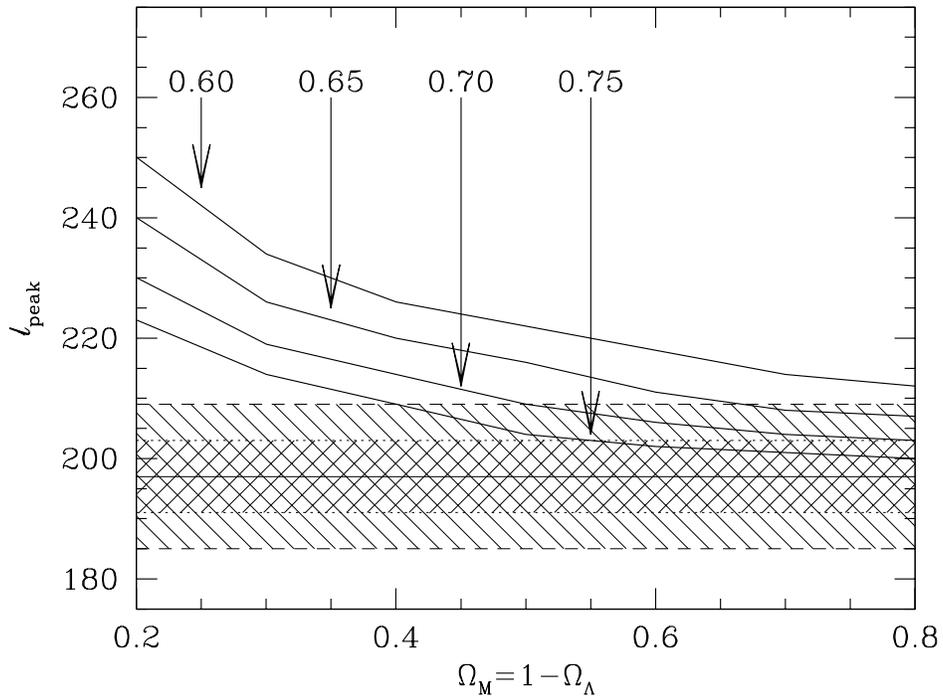}
\end{center}
\caption{The position of the first acoustic peak, $\ell_{\rm peak}$, as
a function of $\Omega_{\rm M}$ in a {\bf flat\/} universe.  In all cases
we have held $\Omega_Bh^2=0.02$.  We show 4 values of the Hubble constant
$H_0=100\,h\,{\rm km}\,{\rm s}^{-1}\,{\rm Mpc}^{-1}$: $h=0.6$, 0.65, 0.70
and 0.75.  The region allowed by the {\sl Boomerang\/} data is shown hatched.}
\label{fig:peak}
\end{figure}

The new {\sl Boomerang\/} power spectrum indicates at first sight a rather
weak second peak.  The important thing to say here is that there is still a
great deal of power at scales $\ell=400$--600; the power spectrum has
certainly not damped to zero.  Guided by the physics of acoustic oscillations,
we will assume that there {\it is\/} a second peak and phrase the question as:
how high is it?
At face value, the data would appear to indicate that any second peak is
rather flat.
However, we would caution that the data are probably more uncertain at
these small scales, since a number of corrections need to be applied.
Uncertainties in beam size, beam asymmetry, pixelization, the effects of
bolometer time constant etc., can all lead to systematic effects on these
scales.  Nevertheless, the {\sl Boomerang\/} team has modeled these effects
and it seems that the basic prediction of the standard $\Lambda$-dominated
model (solid line in Fig.~\ref{fig:bp}), for example, would give a higher
peak than the data seem to indicate.
It is therefore worth investigating how one obtains a lower second peak,
in relation to the first.

A wealth of information is stored in the peak heights, but their signature
is more model dependent than the locations.  To obtain a large ratio of power
between the first and second peaks, we would naturally like to have a large
first peak, which points indirectly to a low matter density
universe.  The argument goes as follows (Hu \& White~\cite{Signatures}).
Two primary effects govern the peak heights: baryon drag and the driving force
of photon self-gravity.
As a perturbation enters the horizon, the fluid is compressed by its
self-gravity.  Photon pressure resists the compression, causing the
photon-baryon contribution to the potential to decay.  The fluid is then
released into the acoustic phase in this highly compressed state.
The photons are intrinsically `hot' and do not need to battle against a
large gravitational potential when streaming to the observer, leading to a
large temperature anisotropy on the sound horizon scale.
This `driving effect' does not occur if the potentials are dominated by
an external source, such as cold dark matter.  Thus the first peak is boosted
relative to the low-$\ell$ plateau in universes with low matter density.
The baryons additionally provide inertia to the photon-baryon fluid, enhancing
the compressions into the potential wells and retarding the rarefactions
struggling against the potentials.  Since the first peak in adiabatic models
is a compression, a high baryon content enhances the first peak.

In addition to lowering the matter density, one could obtain decaying
potentials at last scattering by increasing the radiation content of the
Universe, for example through decaying neutrino models
(Bardeen, Bond \& Efstathiou~\cite{BBE}; Dodelson, Gyuk \& Turner~\cite{DGT})
or volatile neutrino models (Pierpaoli \& Bonometto~\cite{PieBon}).
These models also have an enhanced first peak
(White, Gelmini \& Silk~\cite{WGS}).  For a wide range of neutrino mass
and lifetime, the neutrino decay happens while
the modes relevant to CMB anisotropies are outside the horizon, and thus
decaying neutrino models mimic models with a very large `equivalent number
of neutrinos' $N_\nu$.  As an example, let us consider a model with
$\Omega_{\rm M}\,{=}\,1$ and $h\,{=}\,0.65$.
To fit large-scale structure we want to
increase the horizon at equality by a factor ${\sim}\,2.6$, requiring
$N_\nu\,{\sim}\,44$.  This can be achieved with
$\left({m_\nu/{\rm keV} }\right)^2 \left({\tau/{\rm yr}}\right)\,{\sim}\,500$.
We show the effect of this in Fig.~\ref{fig:bp}.  Note that raising the
radiation density has moved the peaks slightly rightwards -- a change in the
spatial curvature would be necessary to move them left again (e.g.~adding
$\Omega_\Lambda\,{=}\,0.5$, long-dashed line in Fig.~\ref{fig:bp},
fits the data).

Other ways of increasing the effective number of relativistic degrees of
freedom can achieve the same effect of increasing the height of the first
peak compared with the second, for example by adding extra sterile neutrino
species or using large lepton asymmetry (Kinney \& Riotto~\cite{KinRio}).
Note that the addition of massive neutrinos (hot dark matter) will generally
{\it increase\/} the second peak height relative to the first --
which does not help -- although it will move the peaks slightly to the left.
For reasonable neutrino masses this is likely to be a very small effect
however.

The ratio of the heights of the first and second peaks is set by the physics
at recombination and the primordial power spectrum.  Thus it depends on
$\Omega_{\rm M}h^2$, $\Omega_{\rm B}h^2$, $n$ and the radiation energy density
(parameterized by $N_\nu$).  The effect of tilt is to change the ratios by
$(\ell_2/\ell_1)^{n-1}$, we show the effect of changing $\Omega_{\rm M}h^2$
and $\Omega_{\rm B}h^2$ in Table~\ref{tab:ratio}.  The {\sl Boomerang\/}
data give this ratio as approximately 3, so one can see that rather extreme
values of the parameters may be required.

A high baryon fraction, coupled with a low matter density or a high radiation
density, in a model with less small scale power than scale-invariance predicts,
would naturally produce a diminishing series of peaks, including a small
second peak.
Models with less small-scale power than scale-invariance can arise naturally
in inflation ranging from models with power law spectra slightly tilted away
{}from scale-invariance to models with broken power laws or even rapid drops
in power at some scale (see Lyth \& Riotto~\cite{LytRio} for a recent review).
The most natural such models are the tilted models with `red' spectra.
The low matter density and high baryon density help to boost the first peak
enough to overcome some of the effects of the tilt, making the ratio of the
first to second peaks larger.
And, finally, a high baryon density reduces the rarefaction peaks, of which the
second peak is the first example.
We should also point out that low $\Omega_{\rm M}$ models with some tilt and
high $\Omega_{\rm B}$ have their first peak shifted a little to the left
compared with more standard models, though this is a small effect.
Models with red spectra sometimes predict tensor anisotropies, which serve to
lower the whole peak structure relative to the {\sl COBE\/} normalization.
This generally makes it more difficult to obtain the necessary power at the
first peak.

If we tilt the spectrum to remove small scale power we are limited in how
much other small-scale power reducing effects can operate.  Thus a high first
peak in a tilted model limits the epoch of reionization.  Currently limits on
the reionization optical depth are $\tau\la 0.3$
(Griffiths, Barbosa \& Liddle~\cite{GriBL}), although this is somewhat model
dependent.  A strong constraint on $\tau$ requires combining the CMB data with
information from large-scale structure.

More speculatively, the small second peak could be telling us that the
peaks are more `washed out' than the inflationary predictions.
In other words, there may be some decaying mode left in the fluctuations,
or the perturbations may not be entirely synchronized at horizon crossing,
so that there may be some loss of coherence of the oscillations.
Note that the first peak is well defined, so we would require a source which
turned off before those modes entered the horizon, i.e.~was acting only at
early times, perhaps before equality.  Such a source would not be `scaling'
and a mechanism would be required to pick out a preferred scale in the
Universe, e.g.~matter-radiation equality.  Although this does not seem
{\it a priori\/} very likely,
it would be intriguing if the structure of the peaks told us something
fundamental about the origin of the seed perturbations themselves!

\begin{figure}
\begin{center}
\leavevmode
\epsfysize=10cm \epsfbox{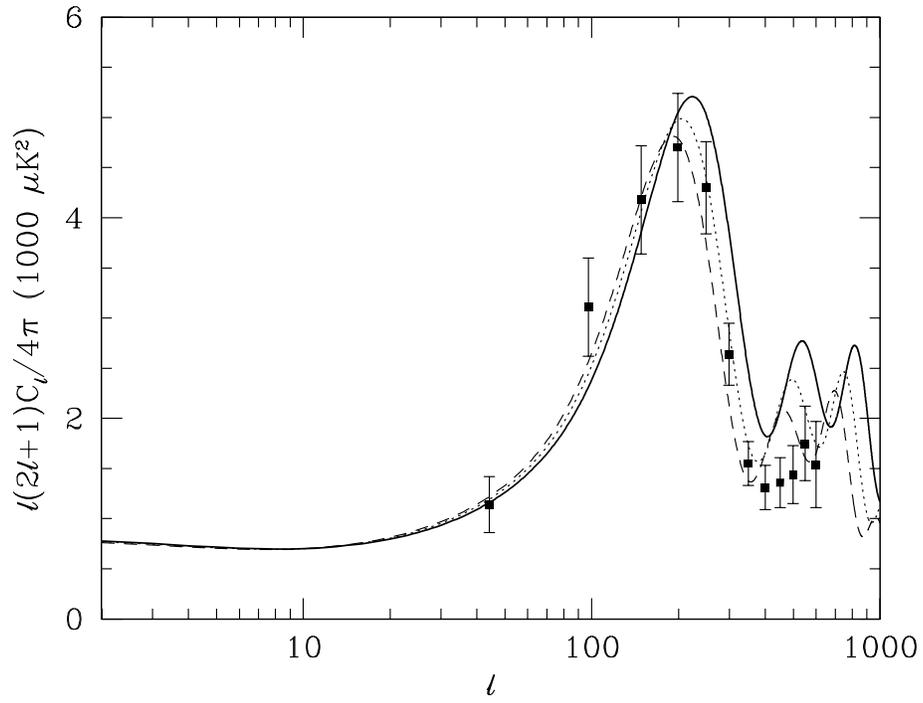}
\end{center}
\caption{The effect of modifying recombination.  Here we have scaled the
energy levels in hydrogen by respectively 10\% (dotted) and 20\% (dashed)
to affect the time of recombination, as a simple way of showing the effect
of bringing the damping tail to lower $\ell$.}
\label{fig:recomb}
\end{figure}

A host of other possibilities exist which seem even less likely.
A source of energy
injection at $z\,{\sim}\,10^3$ could delay recombination and change the damping
of the anisotropies, though one would need to be careful not to distort the
spectrum.  Other changes in the physics of recombination could also increase
damping, although it is hard to believe there is much missing in our
understanding of the physics of hydrogen and helium atoms
(Seager, Sasselov \& Scott~\cite{SSS}).
The damping scale is an integral over the visibility function
(e.g.~Hu \& White~\cite{Damp}), so to move this to lower $\ell$ means
delaying recombination.  We show the kind of effect that would be required in
Fig.~\ref{fig:recomb}, where we reduce the Rydberg energy by 10\% and 20\%.
This delays recombination until lower redshift, mimicking the effect of energy
injection at $z\,{\sim}\,1000$.
Increasing the coupling of the photons to the baryons at higher $z$, for
example by increasing $n_{\rm He}/n_{\rm H}$, has effects that are at the
per cent level, and therefore cannot be significant.
Variation in fundamental physics, such as a changing fine structure constant
(e.g.~Kaplinghat, Scherrer \& Turner~\cite{KapST}) is a more speculative
way to achieve this same goal.  Magnetic fields are often invoked to explain
unexpected phenomena, but here the simplest ideas tend to increase the
small-scale anisotropies.

\section{Conclusions}

The {\sl Boomerang\/} data provide a stunning confirmation of the reality of
acoustic oscillations in the photon-baryon fluid at last scattering.  The fact
that the peak is at $\ell\,{\sim}\,200$ argues that the Universe is close to
spatially flat.  The fact that the second peak appears to be smaller than
naively expected, while explicable within standard models, could be a clue
to something novel in our model of structure formation.

We have argued that the high first peak relative to the second is suggestive
of tilt in the primordial power spectrum, a late epoch of matter-radiation
equality and a
low redshift of reionization.  The slightly leftwards position (relative to the
precisely flat expectation) of the first peak argues
for a short distance to last scattering, and in combination with the former
this argues that the Universe may be (marginally) spatially closed.
Whether the best-fitting models are significantly closed will require
more high precision data.  But in any case, it is now clear that closed models
need to be considered on at least an equal footing with open models when
searching the cosmological parameter space.

The key to making further progress will be the detection of a third peak.
Models with a high baryon content will have a high third peak, tilted models
will have a lower third peak.  Lack of coherence in the oscillations would
be more exciting still, since this would be harder to explain.
The detection of a second feature in the power spectrum would pin down the
fundamental mode of the baryon-photon fluid at last scattering and put us
well on our way towards reconstructing the model of structure formation.

Further measurement of the second peak should come with analysis of the
full {\sl Boomerang\/} 98 data-set, together with data from the
{\sl MAXIMA}\footnote{{\tt http://cfpa.berkeley.edu/group/cmb/}\quad
After this paper was submitted the results of the {\sl MAXIMA}-1
flight were released (Hanany~\cite{Hanany}).
They show impressive confirmation of
the basic picture presented by the {\sl Boomerang\/} data, including the
weakness of the second peak.  Although the {\sl MAXIMA}-1 data show no strong
preference for closed models, we wish to emphasize that there is still a large
region of closed model parameter space which needs to be explored.}
{\sl VSA}\footnote{{\tt http://www.mrao.cam.ac.uk/telescopes/vsa/index.html}},
{\sl DASI}\footnote{{\tt http://astro.uchicago.edu/dasi/}} and
{\sl CBI}\footnote{{\tt http://astro.caltech.edu/$\sim$tjp/CBI/}}
experiments.
In addition long-duration CMB balloon flights in the next couple of seasons,
as well as the imminent launch of
{\sl MAP\/}\footnote{{\tt http://map.gsfc.nasa.gov/}},
should produce much more precise measurements of the relevant $\ell$ range.

The new {\sl Boomerang\/} results have shown a remarkable confirmation of the
conventional picture for structure formation.  On top of that, it is exciting
that the data show some hints of a couple of surprises.  To make it easier to
fit the first peak position, it may be worth bearing in mind the possibility
that the Universe may be spatially closed.
And, for consistency with the structure of the subsidiary peaks, it is worth
keeping an open mind to the possibility that there may yet be some important
physical effects which are not contained within the simplest versions of the
current standard paradigm.

\bigskip
\acknowledgments  
M.~White is supported by the US National Science Foundation,
DS and EP by the Canadian Natural Sciences and Engineering Research Council.
EP is a National Fellow of the Canadian Institute for Theoretical
Astrophysics.

\end{document}